\documentclass[final,3p,times, sort&compress, onecolumn]{elsarticle}
\journal{Entropy}

\usepackage{amsmath}
\begin{document}

\begin{frontmatter}
\title{Ground State, Magnetization Process and  Bipartite Quantum Entanglement \\ of a Spin-1/2 Ising--Heisenberg Model on Planar Lattices of Interconnected Trigonal Bipyramids}
   
\author[1]{Lucia G\'alisov\'a\,\corref{cor1}}
\cortext[cor1]{Corresponding author}
\ead{lucia.galisova@tuke.sk}
\author[2,3]{Michał Kaczor}
\address[1]{
Institute of Manufacturing Management, Faculty of Manufacturing Technologies with the Seat in Pre\v{s}ov, Technical University of~Ko\v{s}ice,\\ Bayerova 1, 080\,01 Pre\v{s}ov, Slovakia}
\address[2]{
The Doctoral School of University of Rzesz\'{o}w, University of Rzesz\'{o}w, Rejtana 16C, 35-935 Rzesz\'{o}w, Poland}
\address[3]{
Insitute of Physics, College of Natural Sciences, University of Rzesz\'{o}w, Rejtana 16A, 35-935 Rzesz\'{o}w, Poland}

\begin{abstract}
The ground state, magnetization scenario and the local bipartite quantum entanglement of a mixed spin-$1/2$ Ising--Heisenberg model in a magnetic field on planar lattices formed by identical corner-sharing bipyramidal plaquettes is  examined by combining the exact analytical concept of generalized decoration-iteration mapping transformations with Monte Carlo simulations utilizing the Metropolis algorithm. The ground-state phase diagram of the model involves six different phases, namely, the standard ferrimagnetic phase, fully saturated phase, two unique quantum ferrimagnetic phases, and two macroscopically degenerate quantum ferrimagnetic phases with two chiral degrees of freedom of the Heisenberg triangular clusters. The diversity of ground-state spin arrangement is manifested themselves in seven different magnetization scenarios with one, two or three fractional plateaus whose values are determined by the number of corner-sharing plaquettes. The low-temperature values of the concurrence demonstrate that the bipartite quantum entanglement of the Heisenberg spins in quantum ferrimagnetic phases is field independent, but twice as strong if the Heisenberg spin arrangement is unique as it is two-fold degenerate.
\end{abstract}

\end{frontmatter}

\section{Introduction}
\label{sec1}

Quantum entanglement has been attracting a lot of attention in the last few years mainly due to its crucial role in the development of quantum computers, superdense coding, quantum communication, quantum teleportation, as well as quantum information theory~\cite{Hor09,Nie10,Jae18}. The application potential of this unique phenomenon also exceeds into the quantum biology~\cite{Bal11,Hue13} and quantum metrology~\cite{Gio11,Fro11}.

In quantum theory, quantum entanglement provides a novel platform for exploring long-range quantum correlations, quantum phase transitions as well as exotic properties of many-body systems~\cite{Osb02,Ost02,Vid03,Ros04}. The low-dimensional Heisenberg spin models, involving quantum fluctuations between spins, play a significant role in this regard because they have been proven to be ideal candidates for a rigorous investigation of the entangled states under the influence of the external stimuli such as magnetic field (homogeneous or inhomogeneous) and/or temperature~\cite{Aso04,Sun06,Can06,Jaf11,Zid14,Ale16,Gha21,Var21,Ros05,Zho15,Nar18,Sha07,Sad21}. Moreover, many analytical and numerical calculations have been performed to examine the tuning of the quantum and thermal bipartite entanglement by varying the exchange anisotropy parameter~\cite{Can06,Gu05,Ros05,Jaf11,Zid14,Zho16,Nar18,Fum21,Sad21,Sad12b}, the uniaxial single-ion anisotropy~\cite{Gha21,Var21}, the Dzyaloshinskii--Moriya interaction (spin-orbit coupling)~\cite{Jaf11,Zid14,Zho15,Zho16,Fum21}, the next-nearest-neighbour interaction~\cite{Sun06,Sha07,Lei19}, as well as by introducing impurities into the system~\cite{Sad12a,Sad12b}.

However, the rigorous investigation of the bipartite entanglement in the pure Heisenberg models represents a complex task, which is considerably limited due to a non-commutability of spin operators in the Hamiltonian. This computational problem makes the rigorous study of the phenomenon in general inaccessible across whole parameter space of the systems. On the other hand, replacing some of the Heisenberg spins with three spin components by the Ising ones with only one ($z$-) component at the nodal lattice sites is the alternative way to exactly examine the entanglement in various simpler mixed-spin Ising--Heisenberg models by using the standard transfer-matrix method~\cite{Bax82} and/or the concept of generalized mapping transformations~\cite{Fis59,Syo72,Roj09,Str10}. 
Taking into account the fact that the finite Heisenberg clusters formed by three-component Heisenberg spins are indirectly coupled with each other through the intermediate one-component Ising spin(s), one finds that the eigenstates of two adjacent Heisenberg clusters are separable. Thus, any quantity measuring the local bipartite entanglement in the considered mixed-spin model can be rigorously calculated for each quantum Heisenberg cluster separately. 

To date, the bipartite entanglement has been rigorously examined in several one- (1D) and two-dimensional (2D) mixed-spin Ising--Heisenberg models formed by the identical Heisenberg dimers or triangular clusters which interact with each other via the intermediate nodal Ising spin(s)~\cite{Ana12,Tor14,Roj17a,Roj17b,Sou19,Kar19,Gal20a,Gal21,Eki20,Gal20b}. The investigations brought a deeper insight into the thermal and magnetic-field-driven changes of the phenomenon~\cite{Ana12, Tor14,Roj17a,Roj17b,Sou19,Kar19,Gal20a,Gal21}, the impact of the model's parameters on the phenomenon~\cite{Ana12,Roj17a,Roj17b,Kar19,Gal20a,Eki20,Gal20b,Gal21}, as well as the evolution of the phenomenon near and above second-order (continuous) phase transitions~\cite{Eki20,Gal20b} without any artefacts arising from approximations. Despite their simplicity and the general opinion that the simpler mixed-spin Ising--Heisenberg systems involving isolated local quantum correlations are artificial models, some of the results were in a very good correspondence with ones obtained for more complex Heisenberg counterparts~\cite{Kar19} and also with experiments~\cite{Sou19,Tor18,Str20}.

In the present paper, we will rigorously solve a spin-$1/2$ Ising--Heisenberg model in a longitudinal magnetic field on 2D lattices formed by identical corner-sharing trigonal bipyramidal plaquettes. Our recent studies~\cite{Gal19,Gal20b} of the model without magnetic field on the particular lattice with four inter-connected bipyramidal units have shown that
this quantum mixed-spin model represents a suitable playground for a rigorous study of various unconventional physical phenomena such as the macroscopic degeneracy of the spontaneous long-range order caused by chiral spin degrees of freedom, the spin frustration, and the bipartite entanglement. The aforementioned findings motivated us to extend the investigation of the model also to the effect of the longitudinal magnetic field. The goals of the present paper are to shed a light on the nature of ground states invoked by the applied field, to identify the actual fractional plateaus in the zero-temperature magnetization process, to find out a general formula describing how the values of these plateaus depend on the current number of interconnected bipyramidal plaquettes and, finally, to quantify the bipartite quantum entanglement between the Heisenberg spins in individual ground states.  

In addition to the academic interest, our investigation of the spin-$1/2$ Ising-Heisen\-berg model on 2D lattices formed by interconnected trigonal bipyramids is motivated by the existence of a class of geometrically frustrated structures, namely cobaltates
YBaCo$_4$O$_7$ (Y denotes a rare-earth ion)~\cite{Val02} and anion-radical salts (MDABCO$^{+}$)(C$_{60}^{\bullet -}$) (MDABCO$^{+}$ represents $N$-methyldiazabicyclooctanium cation, C$_{60}^{\bullet -}$ is a radical anion)~\cite{Ots18}, in which one can clearly identify corner-sharing trigonal bipyramidal clusters. Although the mentioned compounds do not represent a precise experimental realization of the magnetic structure proposed in the present paper, we hope that a targeted design of the magnetic material with a magnetic structure of interconnected trigonal bipyramids is feasible. The targeted chemical synthesis involving highly anisotropic spin carriers such Dy$^{3+}$ or Co$^{2+}$ magnetic ions and anion-radical salts could possibly afford desiring 
%Please confirm that the original meaning has been retained
such a quantum mixed-spin system. The findings presented in this paper could serve as a motivation for chemists to achieve this goal.

The outline of the paper is as follows: in Section~\ref{sec2}, a magnetic structure of the investigated model is described and the most important steps of its rigorous treatment combining the analytical and numerical approaches are clarified. In Section~\ref{sec3}, we present the most interesting numerical results for the ground state and the magnetization process of the model. The section also includes an analysis of the bipartite quantum entanglement in the individual ground states. Finally, the summary of the most important findings are presented in Section~\ref{sec4}.
%

%%%%%%%%%%%%%%%%%%%%%%%%%%%%%%%%%%%%%%%%%%

\section{Model and Its Rigorous Treatment}
\label{sec2}

\begin{figure}[b!]
\centering
\includegraphics[scale=0.5]{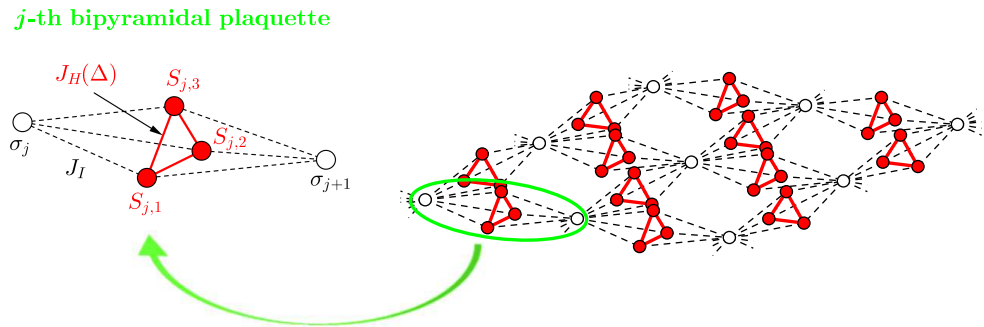}
\caption{A schematic representation of the $j$-th trigonal bipyramidal plaquette and the mixed \mbox{spin-$1/2$} Ising--Heisenberg model on the particular 2D lattice with four ($q=4$) corner-sharing bipyramidal plaquettes. White circles label lattice sites occupied by the Ising spin $\sigma=1/2$ and red circles denote lattice sites occupied by the Heisenberg spin $S=1/2$. Black dashed lines illustrate the Ising-type interaction $J_I$ between the 
Ising and Heisenberg spins and red solid lines indicate XXZ Heisenberg exchange interaction $J_{H}(\Delta)$ between the Heisenberg spins in the plaquette. \label{fig1}
}
\end{figure}

We consider a mixed spin-$1/2$ Ising--Heisenberg model in a longitudinal magnetic field on 2D lattices consisting of identical corner-sharing trigonal bipyramidal plaquettes, as is schematically depicted in Figure~\ref{fig1} for one particular lattice with four such plaquettes. In this figure, the common vertices of plaquettes (white circles) are occupied by the Ising spins $\sigma=1/2$ that interact with other spins solely through their $z$-components. The rest ones (red circles), forming internal equilateral triangles oriented perpendicularly to the plaquette axes, are occupied by the Heisenberg spins $S=1/2$ that are coupled to each other via $x$-, $y$-, and $z$-components. Assuming $q$ bipyramidal plaquettes share a common vertex, the total Hamiltonian of the mixed spin-$1/2$ Ising--Heisenberg model can be written as a sum of 
plaquette (five-spin cluster) Hamiltonians $\hat{\cal{H}} = \sum_{j=1}^{Nq/2} \hat{\cal{H}}_j$, where $N$ labels the total number of the nodal lattice sites occupied by the Ising spins (we consider the thermodynamic limit $N\to \infty$). 
Each plaquette Hamiltonian $\hat{\cal{H}}_j$ contains all exchange interactions realized within the $j$th Ising--Heisenberg trigonal bipyramid and Zeeman terms that describe the influence of the applied external magnetic field on magnetic moments of the individual spins:
\begin{equation}
\label{eq:Hj1}
\begin{aligned}
 \hat{{\cal H}}_j =&
-\,J_H\sum_{k=1}^3\left[\Delta(\hat{S}_{j,k}^x\hat{S}_{j,k+1}^x + \hat{S}_{j,k}^y\hat{S}_{j,k+1}^y)+\hat{S}_{j,k}^z\hat{S}_{j,k+1}^z\right]
-J_{I}\sum_{k=1}^3\hat{S}_{j,k}^{z}(\hat{\sigma}_{j}^z + \hat{\sigma}_{j+1}^z) 
\\
& 
-\,H_{H}\sum_{k=1}^3\hat{S}_{j,k}^{z} - \frac{H_I}{q}(\hat{\sigma}_{j}^z + \hat{\sigma}_{j+1}^z).
\end{aligned}
\end{equation}
%\vspace{-5pt}
\hspace{-0.15cm}
In the above, $\hat{S}_{j,k}^{\alpha}$ ($\alpha = x,y,z$) and $\hat{\sigma}_j^z$ are spatial components of the spin-$1/2$ operator of the Heisenberg spin from the $j$th triangle and $z$-component of the Pauli operator with the eigenvalues $\pm1/2$ at the $j$th nodal lattice site, respectively, which satisfy the periodic boundary conditions $\hat{S}_{j,4}^{\alpha} \equiv \hat{S}_{j,1}^{\alpha}$ and $\hat{\sigma}_{Nq/2+1}^z \equiv \hat{\sigma}_{1}^z$. The parameter $J_H$ marks the XXZ Heisenberg interaction within the Heisenberg triangles, $\Delta$ is the exchange anisotropy parameter in this interaction, and $J_I$ labels the Ising-type interaction between the nearest-neighbouring Ising and Heisenberg spins. The last two terms $H_H$ and $H_I$ in the second line of Equation~(\ref{eq:Hj1}) are Zeeman terms, which account for the magnetostatic energy of the Heisenberg and Ising spins in an applied longitudinal magnetic field, respectively.

As we have shown in our recent work on the zero-field case of the model~\cite{Gal20b}, it is convenient for further calculations to introduce the composite spin operators:
\begin{eqnarray}
\label{eq:ttz}
\hat{\mathbf t}_{j}= \sum_{k=1}^3\hat{\mathbf S}_{j,k}\,,\quad \hat{t}_{j}^{\alpha}= \sum_{k=1}^3\hat{S}_{j,k}^\alpha\,\,\, (\alpha = x,y,z),
\end{eqnarray}
which determine the total spin of the Heisenberg triangular clusters and its spatial components, respectively. 
From the definition of the latter operators, one can easily obtain the spin identity $(\hat{t}_{j}^{\alpha})^2 = 3/4 +2\sum_{k=1}^3\hat{S}_{j,k}^{\alpha}\hat{S}_{j,k+1}^{\alpha}$. This, in combination with the identity for the square of the total composite spin operator $(\hat{\mathbf t}_{j})^2 = \hat{\mathbf t}_{j}\cdot\hat{\mathbf t}_{j}= (\hat{t}_{j}^{x})^2 + (\hat{t}_{j}^{y})^2 + (\hat{t}_{j}^{z})^2$, allows one to find the following two relations for the Heisenberg spin operators from the same triangular cluster: 
\begin{eqnarray}
\label{eq:relationsSxyz}
\sum_{k=1}^3\left(\hat{S}_{j,k}^x\hat{S}_{j,k+1}^x + \hat{S}_{j,k}^y\hat{S}_{j,k+1}^y\right) = \frac{1}{2}\left[(\hat{\mathbf t}_{j})^2 - (\hat{t}_{j}^{z})^2\right] - \frac{3}{4}\,,\quad  \sum_{k=1}^3\hat{S}_{j,k}^z\hat{S}_{j,k+1}^z = \frac{1}{2}(\hat{t}_{j}^{z})^2-\frac{3}{8}.  
\end{eqnarray}

Bearing in mind the above relations and the definition of the $z$-component of the composite spin operator $\hat{t}_{j}^{z}$ listed in Equation~(\ref{eq:ttz}), the plaquette Hamiltonian~(\ref{eq:Hj1}) can be expressed in the alternative form:
\begin{equation}
\label{eq:Hj2}
\begin{aligned}
\hat{{\cal H}}_j =&\,\, \frac{3J_H}{8}(2\Delta\!+\!1) - \frac{J_H\Delta}{2}(\hat{\mathbf t}_{j})^2 + \frac{J_H}{2}(\Delta-1)(\hat{t}_{j}^{z})^2 -J_{I}\hat{t}_{j}^{z}(\hat{\sigma}_{j}^z + \hat{\sigma}_{j+1}^z) \\
  & -\, H_{H}\hat{t}_{j}^{z} - \frac{H_I}{q}(\hat{\sigma}_{j}^z + \hat{\sigma}_{j+1}^z).
\end{aligned}
\end{equation}
It is easy to prove that the operators $(\hat{\mathbf t}_{j})^2$, $\hat{t}_{j}^{z}$ appearing in Equation~(\ref{eq:Hj2}) satisfy the commutation relations $\big[\hat{{\cal H}}_j,(\hat{\mathbf t}_{j})^2\big]=0$ and $\big[\hat{{\cal H}}_j,\hat{t}_{j}^{z}\big]=0$, which implies that they are both conserved quantities with well defined quantum spin numbers $t_j(t_j+1)$ and $t_j^z = \{-t_j, -t_j+1,\ldots,t_j\}$ for $t_j = \{3/2,1/2\}$, respectively. In this regard, \mbox{Equation~(\ref{eq:Hj2})} represents a fully diagonal form of the plaquette Hamiltonian~(\ref{eq:Hj1}), which implies that the corresponding energy eigenvalues can be expressed in terms of the respective quantum spin numbers:
\begin{equation}
\label{eq:Et}
\begin{aligned}
E_{t_j,t_j^z} =&\,\, \frac{3J_H}{8}(2\Delta\!+\!1) - \frac{J_H\Delta}{2}t_j(t_j+1) + \frac{J_H}{2}(\Delta-1)(t_{j}^{z})^2 -J_{I}t_{j}^{z}(\sigma_{j}^z + \sigma_{j+1}^z)\\
 & -\, H_{H}t_{j}^{z} - \frac{H_I}{q}(\sigma_{j}^z + \sigma_{j+1}^z).
\end{aligned}
\end{equation}

At this calculation stage, the partition function of the considered spin-$1/2$ Ising--Heisen\-berg model can be partially factorized due to commuting character of different plaquette Hamiltonians and written in terms of the eigenvalues~(\ref{eq:Et}) of the plaquette Hamiltonian:
\begin{equation}
\label{eq:Z}
{\cal Z} = \textrm{Tr}\,\textrm{exp}\left(-\beta\hat{{\cal H}}\right)  =  \sum_{\{\sigma_n\!\}}\prod_{j=1}^{Nq/2}\textrm{Tr}_{j}\,\textrm{exp}\left(-\beta\hat{{\cal H}}_j\right) = \sum_{\{\sigma_n\!\}}\prod_{j=1}^{Nq/2}\sum_{t_j,t_j^z}g_{t_j}\,\textrm{exp}\big({-}\beta E_{t_j,t_j^z}\big).
\end{equation}
Here, $\beta = 1/(k_{\rm B}T)$ ($k_{\rm B}$ is the Boltzmann's constant and $T$ is the absolute temperature of the system), and the summation symbol $\sum_{\{\sigma_n\!\}}$ denotes a summation over all possible spin configurations of the Ising spins, the product symbol $\prod_{j=1}^{Nq/2}$ runs over all trigonal bipyramids, and the double summation symbol $\sum_{t_j,t_j^z}$ runs over all possible values of the quantum numbers $t_j$, $t_j^z$ of the composite spins. Finally, $g_{t_j}$ is the degeneracy factor, which takes the value~$1$ for the quantum number $t_j =3/2$ and $2$ for the quantum number $t_j =1/2$. 
After performing double summations over $t_j$ and $t_j^z$, one gains the effective Boltzmann's weight $w(\sigma_j^z,\sigma_{j+1}^z)$, which depends only on the Ising spin states $\sigma_j^z$, $\sigma_{j+1}^z$, and, thus, it can be replaced by a simpler but equivalent expression using the generalized decoration-iteration mapping transformation~\cite{Fis59,Syo72,Roj09,Str10}:
\begin{equation}
\label{eq:DIT}
\begin{aligned}
w(\sigma_j^z, \sigma_{j+1}^z) =& \sum_{t_j,t_j^z}g_{t_j}\,\textrm{exp}\big({-}\beta E_{t_j,t_j^z}\big)  = 2\,\textrm{exp}\left[\frac{\beta H_I}{q}(\sigma_j+\sigma_{j+1})-\frac{\beta J_H}{4}\right] \\
& \times\Bigg\{\left[\textrm{exp}\left(\beta J_H\Delta\right)+ 2\,\textrm{exp}\left({-}\frac{\beta J_H\Delta}{2}\right)\right]\cosh\left[\frac{\beta J_I}{2}(\sigma_{j}^{z} + \sigma_{j+1}^{z})+\frac{\beta H_H}{2}\right]
\\
 & 
\hspace{1cm}+\,{\rm exp}\left(\beta J_H\right)\cosh\left[\frac{3\beta J_I}{2}(\sigma_{j}^{z} + \sigma_{j+1}^{z}) + \frac{3\beta H_H}{2}\right]\Bigg\}  \\
=&\, A\,{\rm exp}\left[\beta J_{\rm eff}\,\sigma_j^z\sigma_{j+1}^z + \frac{\beta H_{\rm eff}}{q}(\sigma_{j}^{z} + \sigma_{j+1}^{z})\right].
\end{aligned}
\end{equation}
The novel effective parameters $A$,  $J_{\rm eff}$, and  $H_{\rm eff}$ emerging on the right-hand side of Equation~(\ref{eq:DIT}) are determined by 'self-consistency' of the algebraic approach used:
\begin{equation}
\label{eq:A,Jeff,Heff}
A =
\sqrt[4]{w_{+}w_{-}w_{0}^2}\,, \quad
J_{\rm eff} = k_{\rm B}T\ln\left(\frac{w_{+}w_{-}}{w_{0}^2}\right), \quad
H_{\rm eff} = \frac{k_{\rm B}Tq}{2}\ln\left(\frac{w_{+}}{w_{-}}\right).
\end{equation}
Here, $w_{\pm} = w(\pm1/2, \pm1/2)$ and $w_{0} = w(\pm1/2, \mp1/2)$. 
After substituting Equation~(\ref{eq:DIT}) into Equation~(\ref{eq:Z}), one obtains the rigorous equivalence between the partition function ${\cal Z}$ of the spin-$1/2$ Ising--Heisenberg model given by the Hamiltonian (\ref{eq:Hj1}) and the partition function ${\cal Z}_{IM}$ of the effective spin-$1/2$ Ising model on the corresponding $q$-coordinated 2D lattice given by the Hamiltonian ${\cal H}_{IM}= -J_{\rm eff}\sum_{\langle j,n\rangle}^{Nq/2} \sigma_j^z\sigma_{n}^z - H_{\rm eff}\sum_{j=1}^{N} \sigma_j^z$:
\begin{equation}
\label{eq:ZZI}
{\cal Z}(T, J_H, J_I, \Delta, H_H, H_I) = A^{Nq/2}{\cal Z}_{IM}(T, J_{\rm eff}, H_{\rm eff}).
\end{equation}
The mapping relation~(\ref{eq:ZZI}) represents the crucial result of the rigorous solution of the considered 2D spin-$1/2$ Ising--Heisenberg model in an external magnetic field because of all important physical quantities clarifying a ground-state arrangement, magnetization process, and quantum bipartite entanglement between the Heisenberg spins, namely, the local magnetization $m_I = \langle\hat{\sigma}_{j}^{z}\rangle$ and $m_H=\langle \sum_{k=1}^{3}\hat{S}_{j,k}^{z}\rangle$ per nodal Ising spin and Heisenberg triangular cluster, respectively, the total magnetization $m$ per bipyramidal plaquette, as well as the pair correlation functions $C_{HH}^{xx(yy)} = \langle \hat{S}_{j,k}^{x}\hat{S}_{j,k+1}^{x}\rangle= \langle \hat{S}_{j,k}^{y}\hat{S}_{j,k+1}^{y}\rangle$, $C_{HH}^{zz} = \langle \hat{S}_{j,k}^{z}\hat{S}_{j,k+1}^{z}\rangle$ and $C_{IH}^{zz} = \langle \hat{\sigma}_{j}^{z}\hat{S}_{j,k}^{z}\rangle = \langle \hat{\sigma}_{j+1}^{z}\hat{S}_{j,k}^{z}\rangle$, can be directly derived from the formula for the Gibbs free energy ${\cal G} = -k_{\rm B}T\ln{\cal Z}$ by means of the differential calculus:
\begin{subequations}
\begin{eqnarray}
\label{eq:mI,mH,m}
m_I \!\!\!&=&\!\!\! -\frac{1}{2N}\frac{\partial {\cal G}}{\partial H_I},\quad m_H = -\frac{2}{Nq}\frac{\partial {\cal G}}{\partial H_H},\quad m = \frac{qm_H+2m_I}{q},
\\
\label{eq:Czz,Cxy,CIH}
C_{HH}^{zz} \!\!\!&=&\!\!\!\frac{2}{3NqJ_H}\!\left(\!\Delta\frac{\partial {\cal G}}{\partial \Delta}-\frac{\partial {\cal G}}{\partial J_H}\!\right),\,\,\, C_{HH}^{xx(yy)} = -\frac{1}{3NqJ_H}\frac{\partial {\cal G}}{\partial \Delta},\,\,\,
 C_{IH}^{zz} = -\frac{1}{3Nq}\frac{\partial {\cal G}}{\partial J_I}.
\end{eqnarray}
\end{subequations}
The final analytical expressions of all the physical quantities listed in\linebreak Equations~(\ref{eq:mI,mH,m}) and~(\ref{eq:Czz,Cxy,CIH}) depend on the on-site magnetization $m_{IM}=\langle\sigma_j^z\rangle_{IM}$ and the pair correlation function $C_{IM}^{zz}=\langle\sigma_j^z\sigma_{j+1}^z\rangle_{IM}$ of the effective 2D $q$-coordinated spin-$1/2$ Ising lattice with the temperature-dependent nearest-neighbour interaction $J_{\rm eff}$ in the temperature-dependent magnetic field $H_{\rm eff}$. Because an exact solution for the 2D spin-$1/2$ Ising model in an external magnetic field still belongs to unresolved issues of condensed matter physics, one has to resort to some numerical algorithm applicable to the 2D Ising lattices to gain the accurate results for $m_{IM}$ and $C_{IM}^{zz}$. In the present paper, we will employ the classical Monte Carlo (MC) simulations implementing the standard Metropolis algorithm~\cite{Met53,Met54} for the effective spin-$1/2$ Ising lattice of a sufficiently large linear size $L$. 

%%%%%%%%%%%%%%%%%%%%%%%%%%%%%%%%%%%%%%%%%%
\section{Discussion of the Numerical Results}
\label{sec3}
 
In this section, we will proceed to a discussion of the most interesting numerical results for the 2D spin-$1/2$ Ising--Heisenberg model in an external magnetic field with the antiferromagnetic Ising-type interaction $J_I<0$ between the Ising and Heisenberg spins. For simplicity, we will assume that the local magnetic fields acting in the Ising and Heisenberg spins are identical $H_I=H_H=H$. The absolute value of the interaction $J_I$ will be used as an energy unit for defining a relative strength of the Heisenberg interaction $J_H/|J_I|$ and the magnetic field $H/|J_I|$.
 
\subsection{Ground-State Phase Diagrams}
\label{subsec3.1}

\begin{figure}[h!]
\centering
\includegraphics[scale=0.5]{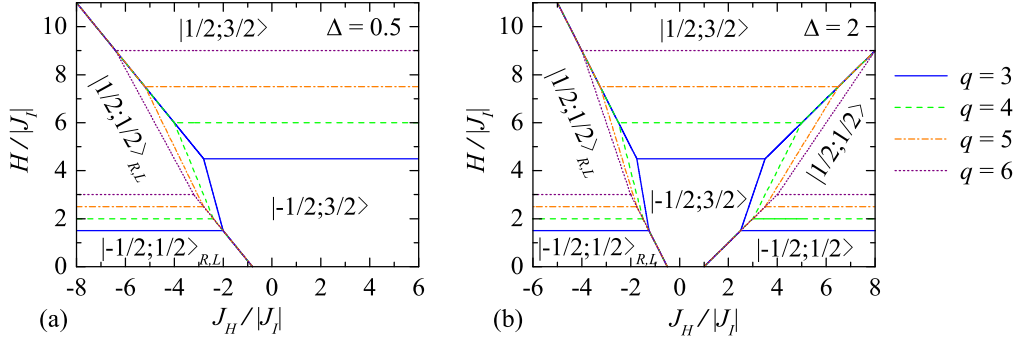}
\caption{The ground-state phase diagram of the spin-$1/2$ Ising--Heisenberg model on 2D lattices with three ($q=3$), four ($q=4$), five ($q=5$), and six ($q=6$) corner-sharing trigonal bipyramidal plaquettes in the $J_H/|J_I| - H/|J_I|$ parameter plane for two representative values of the exchange anisotropy parameter: (\textbf{a})~$\Delta = 0.5$ and (\textbf{b})~$\Delta =2$. \label{fig2}
}
\end{figure}

First, we take a look at possible magnetic ground-state arrangement of the model, which can be determined by a systematic inspection of the eigenvalues~(\ref{eq:Et}) of the plaquette Hamiltonian~(\ref{eq:Hj1}) for all possible combinations of quantum spin numbers $t_j$, $t_j^z$ entering therein. The typical ground-state phase diagrams are depicted in Figure~\ref{fig2} in the\linebreak $J_H/|J_I|-H/|J_I|$ parameter plane for two representative values of the exchange anisotropy $\Delta=0.5$ and $2$ by assuming four different numbers $q$ of corner-sharing trigonal bipyramidal plaquettes forming 2D lattices. As one can see from Figure~\ref{fig2}a, the ground-state phase diagram of the model with the easy-axis exchange anisotropy $\Delta=0.5$ contains four different ground states. Specifically, two ground states are macroscopically degenerate quantum ferrimagnetic phases $|-1/2;1/2\rangle_{R,L}$ and $|1/2;1/2\rangle_{R,L}$, which differ from each other only by the orientation of Ising spins with respect to the applied magnetic field as indicated by the corresponding eigenvectors and eigenenergies per plaquette:
\begin{subequations}
\begin{eqnarray}
\label{eq:pm1/2,1/2 R,L}
|{\pm}1/2;1/2\rangle_{R,L} \!\!\!\!&=&\!\!\!\! \!\prod_{j=1}^{Nq/2}\!
        \left|\pm\right\rangle_{\sigma_j^z}\otimes\left|1/2, R \textrm{ or } L\right\rangle_{\triangle_j},\hspace{0.5cm}\\
\label{eq:E_pm1/2,1/2 R,L}        
E_{|{\pm}1/2;1/2\rangle_{R,L}} \!\!\!\!&=&\!\!\!\! \frac{J_H}{4} + \frac{J_H\Delta}{2} \mp \frac{J_I}{2} - \frac{(q\pm2)H}{2q}.
\end{eqnarray}
\end{subequations}
The state vector $\left|1/2, R \textrm{ or } L\right\rangle_{\triangle_j}$ in Equation~(\ref{eq:pm1/2,1/2 R,L}) describes a quantum superposition of three different up-up-down spin states  of the $j$-th Heisenberg triangular cluster with two opposite ($R$ight- and $L$eft-hand side) chiral degrees of freedom:
\begin{eqnarray}
\label{eq:L,R}
\begin{aligned}
\left|1/2, R\right\rangle_{\triangle_j} \!&= \frac{1}{\sqrt{3}}\!\left(\left|\uparrow\uparrow\downarrow\right\rangle+{\rm e}^{\frac{2\pi{\rm i}}{3}}\left|\uparrow\downarrow\uparrow\right\rangle+{\rm e}^{\frac{4\pi{\rm i}}{3}}\!\left|\downarrow\uparrow\uparrow\right\rangle\right)_{\!\triangle_j},
\\
\left|1/2, L\right\rangle_{\triangle_j} \!&= \frac{1}{\sqrt{3}}\!\left(\left|\uparrow\uparrow\downarrow\right\rangle+{\rm e}^{\frac{4\pi{\rm i}}{3}}\left|\uparrow\downarrow\uparrow\right\rangle+{\rm e}^{\frac{2\pi{\rm i}}{3}}\left|\downarrow\uparrow\uparrow\right\rangle\right)_{\!\triangle_j},
\end{aligned}
\end{eqnarray}
The two-fold degeneracy of each Heisenberg triangle results in the field-independent macroscopic degeneracy
$2^{Nq/2}$ of the phases $|{-}1/2;1/2\rangle_{R,L}$ and $|1/2;1/2\rangle_{R,L}$, which is obviously highly sensitive to the current number $q$ of  interconnected trigonal bipyramidal plaquettes (Heisenberg triangular clusters). The direct relation between the number of plaquettes sharing a common vertex and the macroscopic degeneracy of the phases $|{-}1/2;1/2\rangle_{R,L}$, $|1/2;1/2\rangle_{R,L}$ is also reflected in a current value of the residual entropy per nodal Ising spin observed in both the phases. Specifically, it proportionally grows with~$q$~\cite{Der04}: 
\begin{eqnarray}
\label{eq:Sres}
\frac{{\cal S}_{\rm res}}{Nk_{\rm B}} = \lim_{N\to \infty}\frac{1}{N}\ln 2^{Nq/2}\approx 0.347q.
\end{eqnarray}

The other two ground states are the classical ferrimagnetic phase $|{-}1/2;3/2\rangle$ and the fully saturated phase $|1/2;3/2\rangle$. These two phases again differ from each other only by the orientation of the Ising spins with respect to the applied magnetic field, while the Heisenberg spins are fully polarized into the magnetic field direction without any quantum correlations between their $x$- and $y$-components in both phases:
\begin{subequations}
\begin{eqnarray}
\label{eq:pm1/2,3/2}
|{\pm}1/2;3/2\rangle \!\!\!\!&=&\!\!\!\! \!\prod_{j=1}^{Nq/2}\!
        \left|\pm\right\rangle_{\sigma_j^z}\otimes\left|\uparrow\uparrow\uparrow\right\rangle_{\triangle_j},\\
E_{|{\pm}1/2;3/2\rangle} \!\!\!\!&=&\!\!\!\! -\frac{3J_H}{4} \mp \frac{3J_I}{2} - \frac{(3q\pm2)H}{2q}.
\end{eqnarray}
\end{subequations}
The uniqueness of the classical spin arrangements in the phases $|{-}1/2;3/2\rangle$ and $|1/2;3/2\rangle$ given by Equation~(\ref{eq:pm1/2,3/2}) is reflected in the zero entropy per Ising spin ${\cal S}/(Nk_{\rm B})=0$ in parameter regions corresponding to these phases.

It is obvious from Figure~\ref{fig2}a that the classical ferrimagnetic phase $|{-}1/2;3/2\rangle$ can be detected in the whole parameter region with the ferromagnetic Heisenberg coupling $J_H/|J_I|>0$ and partially also in the region with the antiferromagnetic Heisenberg interaction $J_H/|J_I|<0$. By contrast, the macroscopically degenerate quantum phases $|{-}1/2;1/2\rangle_{L,R}$ and $|1/2;1/2\rangle_{L,R}$ are stable solely for the antiferromagnetic Heisenberg couplings $J_H/|J_I|<0$. 
Finally, the saturated phase $|1/2;3/2\rangle$ represents the actual ground state at high enough magnetic fields regardless of whether the ferro- or antiferromagnetic Heisenberg interaction $J_H/|J_I|$ is considered.

On the other hand, the ground-state phase diagram corresponding to the model with the easy-plane anisotropy $\Delta=2$ is a little more complex (see Figure~\ref{fig2}b). It contains two more ground states in addition to the previous four, namely, the unique quantum ferrimagnetic phases $|{-}1/2;1/2\rangle$ and $|1/2;1/2\rangle$ with the Heisenberg triangular clusters in a symmetric quantum superposition of three possible up-up-down spin states but an opposite orientation of the Ising spins:
\begin{subequations}
\begin{eqnarray}
\label{eq:pm1/2,1/2}
|{\pm}1/2;1/2\rangle \!\!\!\!&=&\!\!\!\! \!\prod_{j=1}^{Nq/2}\! 
         \left|{\pm}\right\rangle_{\sigma_j^z}\otimes\frac{1}{\sqrt{3}}\left(\left|\uparrow\uparrow\downarrow\right\rangle\!+\!\left|\uparrow\downarrow\uparrow\right\rangle\!+\!\left|\downarrow\uparrow\uparrow\right\rangle\right)_{\triangle_j},    \\
\label{eq:E_pm1/2,1/2}        
E_{|{\pm}1/2;1/2\rangle} \!\!\!\!&=&\!\!\!\! \frac{J_H}{4}- J_H\Delta \mp \frac{J_I}{2} - \frac{(q\pm2)H}{2q}.         
\end{eqnarray}
\end{subequations}
As shown in Figure~\ref{fig2}b, both the quantum ferrimagnetic phases $|{-}1/2;1/2\rangle$ and $|1/2;1/2\rangle$ emerge in the ground-state phase diagram exclusively in the parameter region of the ferromagnetic Heisenberg interaction $J_H/|J_I|>0$. Naturally, the zero entropy per Ising spin ${\cal S}/(Nk_{\rm B})=0$ solely can be detected in their stability regions due to unique quantum spin arrangement given by Equation~(\ref{eq:pm1/2,1/2}). 

In addition to their location in the zero-temperature $J_H/|J_I|-H/|J_I|$ parameter plane, it is also possible to understand from Figure~\ref{fig2} how the individual phases develop depending on the number $q$ of corner-sharing bipyramidal plaquettes. Namely, the quantum ferrimagnetic phases $|{-}1/2;1/2\rangle_{R,L}$, $|{-}1/2;1/2\rangle$ and the classical one $|{-}1/2;3/2\rangle$ are gradually extended to stronger magnetic fields with increasing number $q$ of the corner-haring plaquettes. Moreover, the classical phase $|{-}1/2;3/2\rangle$ simultaneously spreads to the regions of stronger antiferromagnetic (ferromagnetic) Heisenberg interactions $J_H/|J_I|<0$ ($J_H/|J_I|>0$). The remaining three phases $|1/2;1/2\rangle_{R,L}$, $|1/2;1/2\rangle$ and $|1/2;3/2\rangle$ faithfully follow the evolution of the adjacent ones $|{-}1/2;1/2\rangle_{R,L}$, $|{-}1/2;1/2\rangle$, $|{-}1/2;3/2\rangle$: the quantum phases $|1/2;1/2\rangle_{R,L}$, $|1/2;1/2\rangle$ are gradually shifted to stronger antiferromagnetic and ferromagnetic Heisenberg interactions $J_H/|J_I|<0$ and $J_H/|J_I|>0$, respectively, and the saturated one $|1/2;3/2\rangle$ is shifted to stronger magnetic fields.

\subsection{Magnetization Process}
\label{subsec3.2}

The rich ground-state phase diagrams depicted in Figure~\ref{fig2} suggest various zero-temperature magnetization scenarios of the studied spin-$1/2$ Ising--Heisenberg model either with one, two or three different plateaus at fractional values of the saturation magnetization $m_{sat} = (3q+2)/(2q)$.  In accordance with the definition of the total magnetization $m$ per plaquette listed in Equation~(\ref{eq:mI,mH,m}), the values of these plateaus are given by a current number $q$ of the trigonal bipyramids sharing a common vertex: 
\begin{equation}
\label{eq:plateau}
\frac{m}{m_{sat}} = \frac{q-2}{3q+2},\quad \frac{q+2}{3q+2},\quad  \frac{3q-2}{3q+2}.
\end{equation}
The first (lowest) magnetization plateau at $m/m_{sat} = (q-2)/(3q+2)$ can be identified at low magnetic fields in the stability regions of the macroscopically degenerate quantum ferrimagnetic phase $|{-}1/2;1/2\rangle_{R,L}$ and the unique quantum ferrimagnetic phase $|{-}1/2;1/2\rangle$. The second one at $m/m_{sat} = (q+2)/(3q+2)$ is a result of the spin arrangements  present in the quantum phases $|1/2;1/2\rangle_{R,L}$ and $|1/2;1/2\rangle$, and therefore it can be found at moderate magnetic fields. The highest fractional plateau at $m/m_{sat} = (3q-2)/(3q+2)$ relates to the classical ferrimagnetic phase $|{-}1/2;3/2\rangle$.

\begin{figure}[h!]
\vspace{3mm}
\centering
\includegraphics[scale=0.45]{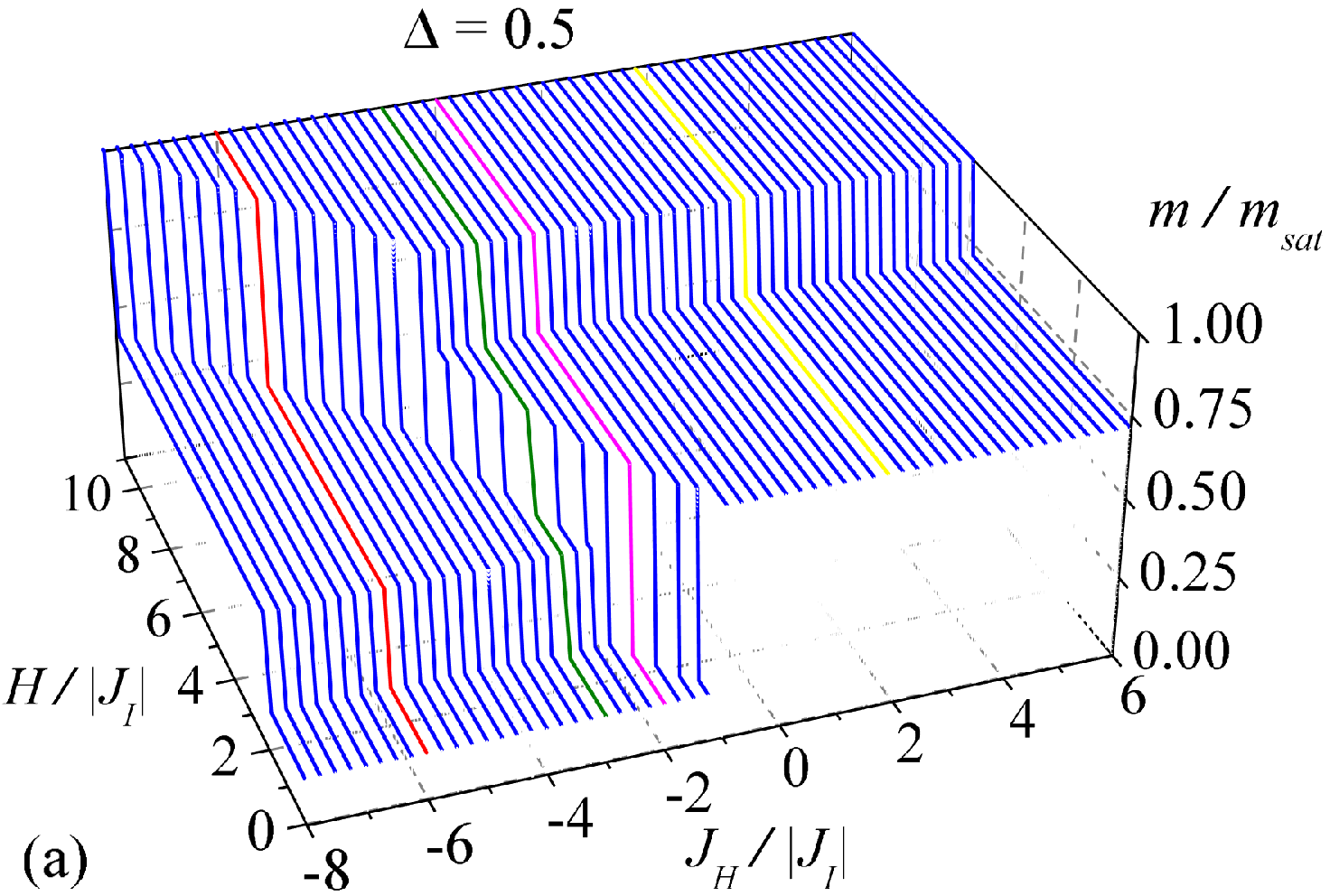}
\hspace{-0.15cm}
\includegraphics[scale=0.45]{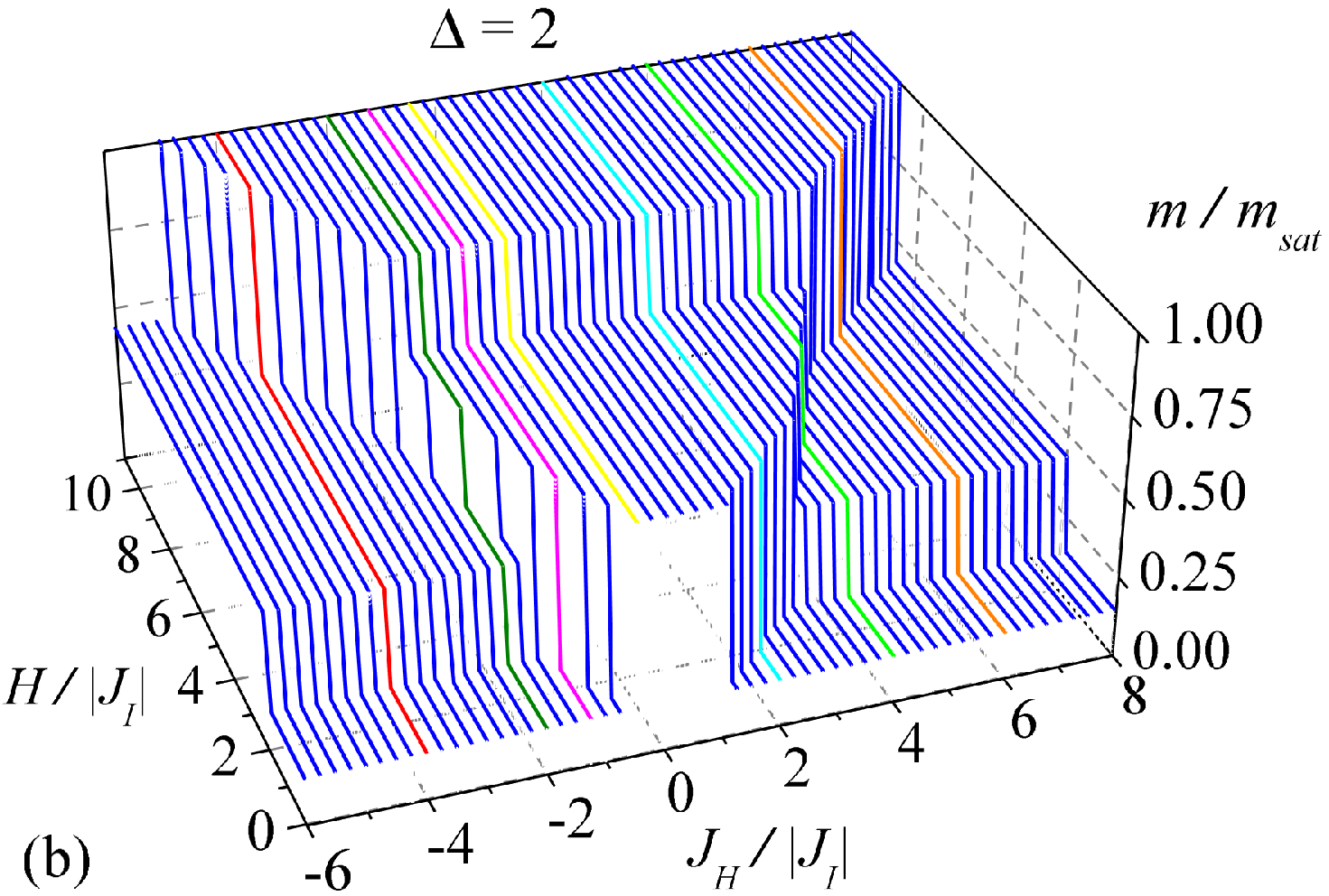}
\\
\includegraphics[scale=0.45]{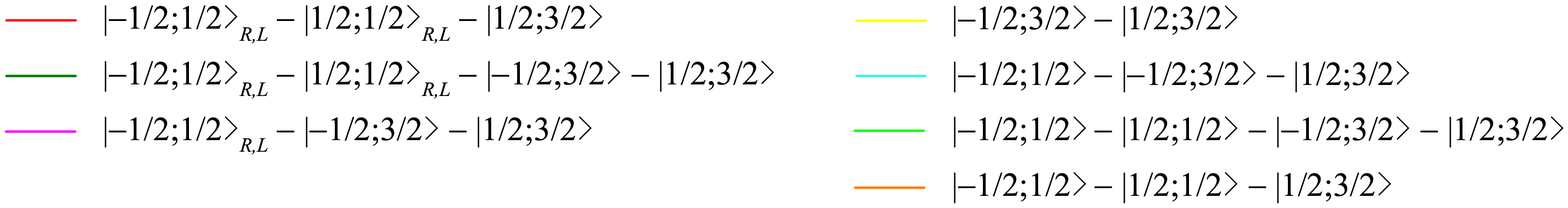}
\caption{3D plots of the total magnetization $m$ of the spin-$1/2$ Ising--Heisenberg model on the regular lattice with four corner-sharing bipyramidal plaquettes reduced to its saturation value $m_{sat}$ as a function of the magnetic field $H/|J_I|$ and the interaction ratio $J_H/|J_I|$ for the exchange anisotropy (\textbf{a})~$\Delta=0.5$ and (\textbf{b})~$\Delta = 2$ at the temperature $k_{\rm B}T/|J_I| = 1.5\times10^{-3}$ obtained by MC simulations for the lattice with $100\times100$ nodal Ising spins (19,800 bipyramidal plaquettes) by using \mbox{$12\times10^4$ MC steps} per node. The curves of distinct colors refer to different magnetization scenarios listed in \mbox{the legend}.\label{fig3}
}
\end{figure}

To illustrate the above statements, two three-dimensional (3D) plots of the isothermal magnetization curves for the particular version of the lattice with four ($q=4$) corner-sharing bipyramidal plaquettes are depicted in Figure~\ref{fig3} at the sufficiently low temperature $k_{\rm B}T/|J_I| = 1.5\times10^{-3}$. The plots are the outcomes of MC simulations for $100\times100$ nodal lattice sites (Ising spins), which corresponds to 19,800 corner-sharing trigonal bipyramidal plaquettes. The adequate numerical accuracy was achieved by $12\times10^4$ MC steps per node. For easy reference, the interaction ratio~$J_H/|J_I|$ is fixed to the same ranges and the anisotropy parameter~$\Delta$ to the same values as were used in Figure~\ref{fig2}. 
It can be understood from a comparison of these plots with corresponding ground-state phase diagrams in Figure~\ref{fig2} that the displayed low-temperature magnetization curves faithfully reflect up to seven different types of zero-temperature magnetization scenarios with the real $1/7$-, $3/7$-, and/or $5/7$-plateaus satisfying the general formulas listed in Equation~(\ref{eq:plateau}):
\begin{itemize}
    \item[i.] $|{-}1/2;1/2\rangle_{R,L}\!-\!|1/2;1/2\rangle_{R,L}\!-\!|1/2;3/2\rangle$,
    \item[ii.] $|{-}1/2;1/2\rangle_{R,L}\!-\!|1/2;1/2\rangle_{R,L}\!-\!|{-}1/2;3/2\rangle\!-\!|1/2;3/2\rangle$,
    \item[iii.] $|{-}1/2;1/2\rangle_{R,L}\!-\!|{-}1/2;3/2\rangle\!-\!|1/2;3/2\rangle$, 
    \item[iv.] $|{-}1/2;3/2\rangle\!-\!|1/2;3/2\rangle$,
    \item[v.] $|{-}1/2;1/2\rangle\!-\!|{-}1/2;3/2\rangle\!-\!|1/2;3/2\rangle$,
    \item[vi.] $|{-}1/2;1/2\rangle\!-\!|1/2;1/2\rangle\!-\!|{-}1/2;3/2\rangle\!-\!|1/2;3/2\rangle$,
    \item[vii.] $|{-}1/2;1/2\rangle\!-\!|1/2;1/2\rangle\!-\!|1/2;3/2\rangle$
\end{itemize}
(see the magnetization curves of different colors). Steep continuous rises between different fractional plateaus as well as between fractional plateaus and the saturation magnetization indicate a presence of the discontinuous magnetization jumps that exist at the critical fields corresponding to the first-order phase transitions only at zero temperature. In agreement with the ground-state analysis performed in Section~\ref{subsec3.1}, the first three magnetization processes i.--iii., which contain the macroscopically degenerate quantum phases $|{\pm}1/2;1/2\rangle_{R,L}$, can be observed for the easy-axis exchange anisotropy $\Delta=0.5$ and also the easy-plane exchange anisotropy $\Delta=2$, but only in the parameter region of the antiferromagnetic Heisenberg interactions $J_H/|J_I|<0$.
On the other hand, the magnetization scenario iv., reflecting the single field-induced transition from the classical ferrimagnetic phase $|{-}1/2;3/2\rangle$ to the saturated one $|1/2;3/2\rangle$, appears for both the antiferromagnetic ($J_H/|J_I|<0$) and ferromagnetic ($J_H/|J_I|>0$) Heisenberg couplings. The last three magnetization processes v.--vii., which involve unique quantum ferrimagnetic phases $|{\pm}1/2;1/2\rangle$, emerge in the parameter region of the ferromagnetic Heisenberg couplings $J_H/|J_I|>0$ under the condition $\Delta>1$.  It should be noted for completeness that the steep staircase dependences of all magnetization curves plotted in Figure~\ref{fig3} are generally gradually smoothing upon increasing of temperature due to a~thermal activation of excited states, until they completely disappear. 
%MDPI:please change hyphen to minus

\subsection{Quantum Bipartite Entanglement}
\label{subsec3.3}

The discussion in the last subsection will be devoted to a bipartite quantum entanglement of the Heisenberg spins in the individual ground states. It is obvious from the plaquette Hamiltonian~(\ref{eq:Hj1}) that the spins may be quantum-mechanically entangled only within the Heisenberg triangular clusters in individual plaquettes. Those from different plaquettes can never be entangled due to the Ising spin at their common vertices.

In general, a degree of the bipartite quantum entanglement between the Heisenberg spins at $k$-th and $(k{+}1)$-th vertex of the $j$-th plaquette can be quantified by the quantity referred to as concurrence~\cite{Woo98}. For the studied 2D spin-$1/2$ Ising--Heisenberg model, the concurrence can be simply calculated from the local magnetization $m_{H}$ of the Heisenberg triangular cluster and the corresponding pair correlation functions $C_{HH}^{xx(yy)}$, $C_{HH}^{zz}$ defined by Equations~(\ref{eq:mI,mH,m}) and (\ref{eq:Czz,Cxy,CIH}) through the following formula~\cite{Ami08,Ost13}:
\begin{equation}
\label{eq:Conc_k,k+1}    
{\cal C}_{k,k+1} = \textrm{max}\left\{0, 4|C_{HH}^{xx(yy)}| - 2\sqrt{\left(\frac{1}{4}+C_{HH}^{zz}\!\right)^{\!2} - \left(\frac{m_{H}}{3}\right)^{\!2}}\right\}.
\end{equation}
Of course, the identical XXZ exchange coupling $J_H(\Delta)$ in a given Heisenberg triangle results in the same degree of the bipartite entanglement of the spin pairs. This is reflected in the same values of the corresponding concurrences:
\begin{equation}
\label{eq:Conc}    
{\cal C}_{1,2} =  {\cal C}_{2,3} = {\cal C}_{3,1} = {\cal C}.
\end{equation}

\begin{figure}[b!]
\centering
\includegraphics[width=0.4\columnwidth]{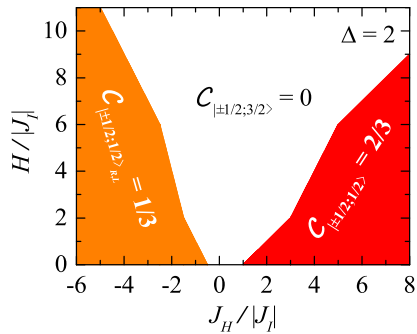}
\caption{The low-temperature ($k_{\rm B}T/|J|=1.5\times10^{-3}$) density map of the concurrence~${\cal C}$ of the spin-$1/2$ Ising--Heisenberg model on the regular lattice with four corner-sharing trigonal bipyramidal plaquettes in the $J_H/|J_I| - H/|J_I|$ plane for the exchange anisotropy parameter $\Delta =2$ constructed from MC simulations performed for the lattice of 19,800 bipyramidal plaquettes by using $2\times10^7$ MC steps per nodal Ising spin. \label{fig4}
}
 %MDPI:please change hyphen to minus
\end{figure}

The global picture on a degree of the bipartite quantum entanglement between the Heisenberg spin pairs in the individual ground-state phases is illustrated in \mbox{Figure~\ref{fig4}}, which shows the low-temperature ($k_{\rm B}T/|J|=1.5\times10^{-3}$) density map of the concurrence~${\cal C}$ of the spin-$1/2$ Ising--Heisenberg model on the regular 2D lattice with four corner-sharing bipyramidal plaquettes in the $J_H/|J_I| - H/|J_I|$ plane for the fixed value of the exchange anisotropy parameter $\Delta=2$. The plotted data have again been obtained by MC simulations performed for the lattice of 19,800 corner-sharing bipyramidal plaquettes, whereas \mbox{$2\times10^7$ MC steps} per nodal Ising spin were used to achieve the accuracy better than $10^{-3}$. It is clear from a direct comparison of Figure~\ref{fig4} with the corresponding ground-state phase diagram depicted in Figure~\ref{fig2}b that the Heisenberg spins forming triangular clusters are quantum-mechanically entangled only if the macroscopically degenerate phases $|{\pm}1/2;1/2\rangle_{R,L}$ and the unique phases $|{\pm}1/2;1/2\rangle$ are ground states. Due to the macroscopic degeneracy caused by two possible chiral degrees of freedom of each triangular cluster, the bipartite entanglement of the Heisenberg spins in former two phases is half weaker than that in latter ones. This is also proven by the corresponding zero-temperature asymptotic values of the concurrence ${\cal C}_{|{\pm}1/2;1/2\rangle_{R,L}} = 1/3$ and ${\cal C}_{|{\pm}1/2;1/2\rangle} = 2/3$. On the other hand, the remaining white region with the zero concurrence ${\cal C}_{|{\pm}1/2;3/2\rangle} = 0$ confirms completely non-entangled arrangements of the Heisenberg spins in the classical ferrimagnetic phase $|{-}1/2;3/2\rangle$ and the fully saturated phase $|1/2;3/2\rangle$.

\begin{figure}[t!]
\centering
\includegraphics[scale=0.5]{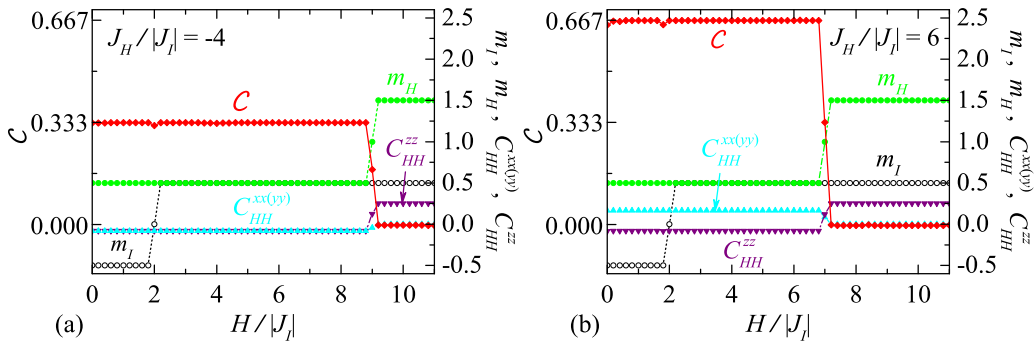}
\caption{The low-temperature ($k_{\rm B}T/|J_I|= 1.5\times10^{-3}$) dependencies of the concurrence ${\cal C}$, the sub-lattice magnetization $m_I$, $m_H$, and the pair correlation functions $C_{HH}^{xx(yy)}$, $C_{HH}^{zz}$ on the magnetic field $H/|J_I|$ of the spin-$1/2$ Ising--Heisenberg model on the regular lattice with four corner-sharing bipyramidal plaquettes for the exchange anisotropy $\Delta = 2$ and two particular interaction ratios (\textbf{a})~$J_H/|J_I| = -4$ and (\textbf{b})~$J_H/|J_I| =6$. The curves are results of the MC simulations for the lattice of $100\times100$ nodal Ising spins by using $2\times10^{7}$ MC steps per node. \label{fig5}
}
\end{figure}

To get a deeper insight onto a role of pair correlations between the Heisenberg spins in their bipartite quantum entanglement, the concurrence ${\cal C}$ as function of the magnetic field $H/|J_I|$ and the corresponding dependencies of the pair correlation functions $C_{HH}^{xx(yy)}$, $C_{HH}^{zz}$ are plotted in Figure~\ref{fig5} for the anisotropy parameter $\Delta =  2$ and two selected interaction ratios $J_{H}/|J_{I}|=-4$ and~$6$ at the temperature $k_{\rm B}T/|J|=1.5\times10^{-3}$. The variations are completed by low-temperature dependences of the local magnetization $m_{I}$ and $m_{H}$ to facilitate identification of the current ground-state spin arrangement.
We note for completeness that all the curves are results of the MC simulations for the lattice of \mbox{$100\times100$ nodal} Ising spins by using $2\times10^{7}$ MC steps per node to achieve accuracy better than $10^{-3}$.

Figure~\ref{fig5}a captures the sequence of field-induced phase transitions  $|{-}1/2;1/2\rangle_{R,L}\!-\!|1/2;1/2\rangle_{R,L}\!-\!|1/2;3/2\rangle$. Evidently, the nonzero concurrence ${\cal C}= 1/3$, which can be found at the magnetic fields $H/|J_{I}|<9$ due to stability of the macroscopically degenerate quantum phases $|{\pm}1/2;1/2\rangle_{R,L}$, is a result of the negative pair correlation functions $C_{HH}^{xx(yy)} = C_{HH}^{zz} = -1/12$ and the reduced local magnetization $m_{H} = 1/2$. Identical values of the transverse and longitudinal correlation functions and their minus sign clearly indicate that the macroscopically degenerate unsaturated bipartite entanglement of the Heisenberg spins from the same XXZ triangle comes from antiferromagnetic $xx (yy)$ correlations of these spins, which are of the same strength and character as those in \mbox{$z$-axis direction}.

A different situation can be found in Figure~\ref{fig5}b, which illustrates the sequence of field-induced phase transitions $|{-}1/2;1/2\rangle\!-\!|1/2;1/2\rangle\!-\!|1/2;3/2\rangle$. Here, the low-temperature concurrence ${\cal C}= 2/3$, which can be observed at the magnetic fields $H/|J_{I}|<7$ due to the presence of the unique quantum phases $|{\pm}1/2;1/2\rangle$, comes from the positive value of the transverse pair correlation function(s) $C_{HH}^{xx(yy)} = 1/6$, the negative longitudinal correlation function $C_{HH}^{zz} = -1/12$, and the reduced local magnetization $m_{H} = 1/2$. It is easy to understand from the values of $C_{HH}^{xx(yy)}$ and  $C_{HH}^{zz}$ that the origin of the local quantum bipartite entanglement of the Heisenberg spins peculiar to the unique quantum phases $|{\pm}1/2;1/2\rangle$ lies in ferromagnetic $xx (yy)$ correlations and these are twice as strong as the antiferromagnetic ones along $z$-axis.

%%%%%%%%%%%%%%%%%%%%%%%%%%%%%%%%%%%%%%%%%%
\section{Conclusions}
\label{sec4}

In the present work, we have comprehensively studied the ground-state properties, possible magnetization scenarios, and the local bipartite quantum entanglement of the Heisenberg spins in the individual quantum ground states of the mixed spin-$1/2$ Ising--Heisenberg model in a longitudinal magnetic field on 2D lattices formed by identical corner-sharing bipyramidal plaquettes. The numerical results have been obtained by combining the exact analytical approach called the decoration-iteration mapping transformation~\cite{Fis59,Syo72,Roj09,Str10} with numerical Monte Carlo simulations utilizing the Metropolis algorithm~\cite{Met53,Met54}.

It has been demonstrated that the ground-state phase diagram of the investigated quantum mixed-spin model qualitatively does not depend on its lattice topology (the number $q$ of corner-sharing plaquettes). In general, it involves in total six different ground states, namely the standard ferrimagnetic phase, fully saturated phase, two unique quantum ferrimagnetic phases and two macroscopically degenerate quantum ferrimagnetic phases with two chiral degrees of freedom of the Heisenberg spins forming triangular clusters in plaquettes. It is also proven that the diversity of the ground-state phase diagram gives rise to seven different magnetization scenarios with one, two or up to three fractional plateaus. The magnitudes of these plateaus are determined by the current number $q$ of the corner-sharing plaquettes.

Other interesting findings are concerned with the bipartite quantum entanglement, which has been  quantified by the concept of the concurrence. We have verified that the Heisenberg spins of the same XXZ triangular cluster of a given plaquette can be entangled either due to stability of the unique quantum ferrimagnetic phases, where they are in a symmetric quantum superposition of three possible up-up-down states, or due to macroscopically degenerate quantum ferrimagnetic phases characterized by two chiral degrees of freedom of each Heisenberg triangle. The strength of the entanglement in all the phases does not depend on the applied magnetic field. Moreover, the corresponding values of concurrence clearly indicate that the entanglement of the Heisenberg spins is twice as strong when their arrangement is unique that when it is two-fold degenerate. Thus, it can be concluded that the macroscopic degeneracy of the Heisenberg triangles proportionally reduces the bipartite quantum entanglement of their spins.

Following our recent paper~\cite{Gal20b}, dealing with the spin-$1/2$ Ising--Heisenberg model on the planar lattice formed by trigonal bipyramids without a magnetic field, there is strong indication that the bipartite entanglement between the Heisenberg spins observed in the unique ferrimagnetic and macroscopically degenerated ferrimagnetic ground-state phases in the present paper will also persist at finite temperatures. Moreover, the interesting thermally induced reentrant behavior of the phenomenon can be expected near first- and second-order phase transitions of the system. Our future investigation will continue in \mbox{this direction}.

%%%%%%%%%%%%%%%%%%%%%%%%%%%%%%%%%%%%%%%%%%
\vspace{6pt} 

%%%%%%%%%%%%%%%%%%%%%%%%%%%%%%%%%%%%%%%%%%
\section*{Aknowledgements}
This work was funded by the Slovak Research and Development Agency under Grant No. APVV-20-0150 and by The Ministry of Education, Science, Research and Sport of the Slovak Republic under Grant No. KEGA 021TUKE-4/2020. M.K. is grateful to the NAWA (Polish National Agency for Academic Exchange) PROM programme for financially supporting the fellowship at the Faculty of Manufacturing Technologies with the seat in Pre\v{s}ov, Technical University of~Ko\v{s}ice in Slovakia.

\end{document}